\documentclass[reprint,showpacs,amsmath,amssymb,pra]{revtex4-1}
\usepackage{graphicx}
\usepackage{dcolumn}
\usepackage{bm}
\usepackage{color}

\begin{document}

\title{Disordered resonant media: Self-induced transparency versus light localization}

\author{Denis~V.~Novitsky$^{1,2}$}

\email{dvnovitsky@gmail.com}

\affiliation{$^1$ITMO University, 49 Kronverksky Prospekt, St.
Petersburg 197101, Russia \\ $^2$B. I. Stepanov Institute of
Physics, National Academy of Sciences of Belarus, 68 Nezavisimosti
Avenue, Minsk 220072, Belarus}

\date{\today}

\begin{abstract}
We propose a concept of disordered resonant media, which are
characterized by random variations of their parameters along the
light propagation direction. In particular, a simple model of
disorder considered in the paper implies random change of the
density of active particles (two-level atoms). Within this model,
the effect of disorder on self-induced transparency (SIT) is
analyzed using numerical simulations of light pulse propagation
through the medium. The transition from the SIT to localization
regime is revealed as well as its dependence on the disorder level,
atom density, medium thickness, and period of random variations.
\end{abstract}

\maketitle

\section{Introduction}

Study of light interaction with disordered media is one of the
actively developing fields of modern optics and photonics
\cite{Wiersma}. Primarily, this is connected with the fundamental
importance and possible applications of the Anderson localization of
light, which was experimentally observed recently \cite{Segev}. The
high prospects of disordered photonics are usually linked to the
development of new scattering structures with the required
characteristics, which can be used as better solar elements, optical
resonators, and laser media.

Introduction of nonlinearity in a disordered medium results in a
whole new area of research with its own problems and relationships.
Light dynamics in nonlinear disordered media are extremely rich, so
that the main problem of ``nonlinearity vs disorder'' is largely
unexplored. We mention here only a few recent results. The
transition from diffusion to localization of light was studied in a
number of different situations, such as three-dimensional disordered
Kerr media \cite{Conti2007}, disordered media with quadratic
\cite{Folli13} and nonlocal nonlinearities \cite{Folli12},
$\mathcal{PT}$-symmetric disordered optical lattices \cite{JovicOL},
and the interface between linear and nonlinear disordered media
\cite{JovicPRA}. The same problem of influence of nonlinearity on
localization is also under thorough investigation in other
frameworks, e.g., in gases of interacting particles
\cite{Cherroret}. Except for localization, a variety of effects
reported includes ``locked explosion'' and diffusive collapse of
wave packets in two-dimensional nonlinear disordered media
\cite{Schwiete}, nonreciprocity due to ultrafast nonlinear dynamics
\cite{Muskens}, persistence of chaotic dynamics \cite{Skokos} and
unusual wave spreading regimes of flat band states in the presence
of nonlinearity \cite{Leykam}, formation of soliton-like states in
random media \cite{Conti2012}, etc. We should also note our previous
papers \cite{Novitsky2014,Novitsky2015}, where self-trapping of
ultrashort pulses in one-dimensional disordered photonic crystals
with relaxing cubic nonlinearity was studied.

Although there are many studies of nonlinear effects in disordered
systems, most of them deal with nonresonant nonlinearities of the
second or third order. Therefore, it seems to be of great interest
to investigate optical response of the disordered media when the
light frequency is close to the frequency of atomic resonance. This
proximity to the resonance results in a bunch of nonlinear optical
effects such as self-induced transparency (SIT)
\cite{McCall,Poluektov}, optical bistability \cite{Hopf,Afan98},
optical kinks \cite{Ponomarenko,Novitsky2017}, unipolar pulse
generation \cite{Pakhomov}, population density gratings formation
\cite{Arkhipov2016,Arkhipov2017}, etc. However, resonant
nonlinearities are rarely considered in the literature on disordered
photonics. An important exception is the work by Folli and Conti
\cite{Folli11}, who proposed the idea of pumping localized Anderson
states by means of SIT pulses in a disordered two-level medium. The
Anderson states are in many ways analogous to the modes of laser
resonators, which allows us to treat this scheme as an unusual
two-level lasing medium.

In this paper, we propose a concept of disordered resonant media and
introduce a simple model of disorder with random variations of the
density of active particles along the direction of light
propagation. This approach is of general interest, since the density
of resonant media is usually assumed to be uniform. Our aim is to
study the influence of this disorder on SIT of light pulses in such
a medium. The paper includes two main sections: Section \ref{eqs} is
devoted to discussion of the methods and parameters used in our
investigation, including the model of disorder. Section \ref{result}
contains the results of numerical simulations and their analysis. A
brief conclusion summarizes the article.

\section{\label{eqs}Main equations and parameters}

\begin{figure}[t!]
\includegraphics[scale=0.65, clip=]{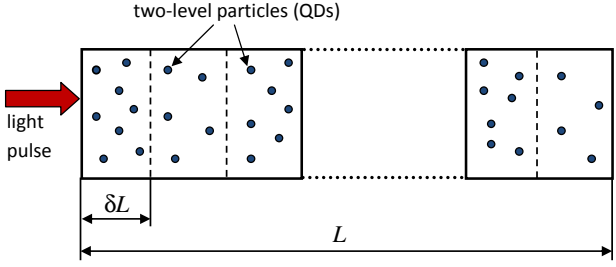}
\caption{\label{fig0} (Color online) The system considered in the
paper: active (two-level) atoms (e.g., quantum dots) dispersed over
the volume of the background dielectric. The density of active
particles randomly changes along the direction of light pulse
propagation with the period $\delta L$.}
\end{figure}

We confine our consideration to a homogeneously broadened two-level
medium and treat it semiclassically with the optical Bloch
equations. The medium (Fig. \ref{fig0}) consists of a background
dielectric doped with active (two-level) atoms, which initially are
in the ground state. Light propagation in this medium is described
by the well-known system of Maxwell-Bloch equations, which includes
differential equations for the dimensionless electric-field
amplitude $\Omega=(\mu/\hbar \omega) E$ (normalized Rabi frequency),
complex amplitude of the atomic polarization $\rho$, and difference
between populations of ground and excited states $w$
\cite{Novit2011}:
\begin{eqnarray}
\frac{d\rho}{d\tau}&=& i l \Omega w + i \rho \delta - \gamma_2 \rho, \label{dPdtau} \\
\frac{dw}{d\tau}&=&2 i (l^* \Omega^* \rho - \rho^* l \Omega) -
\gamma_1 (w-1),
\label{dNdtau} \\
\frac{\partial^2 \Omega}{\partial \xi^2}&-& n_d^2 \frac{\partial^2
\Omega}{\partial \tau^2}+2 i \frac{\partial \Omega}{\partial \xi}+2
i n_d^2 \frac{\partial \Omega}{\partial
\tau} + (n_d^2-1) \Omega \nonumber \\
&&=3 \epsilon l \left(\frac{\partial^2 \rho}{\partial \tau^2}-2 i
\frac{\partial \rho}{\partial \tau}-\rho\right), \label{Maxdl}
\end{eqnarray}
where $\mu$ is the dipole moment of the quantum transition, $\hbar$
is the reduced Planck constant, $\delta=\Delta
\omega/\omega=(\omega_0-\omega)/\omega$ is the normalized frequency
detuning, $\omega$ is the carrier frequency, $\omega_0$ is the
frequency of the atomic resonance, $\gamma_{1}=1/(\omega T_{1})$ and
$\gamma_{2}=1/(\omega T_{2})$ are the normalized relaxation rates of
population and polarization respectively, and $T_1$ ($T_2$) is the
longitudinal (transverse) relaxation time. Light-matter coupling is
described by the dimensionless parameter $\epsilon= \omega_L /
\omega = 4 \pi \mu^2 C/3 \hbar \omega$, where $C$ is the density
(concentration) of two-level atoms and $\omega_L$ is the normalized
Lorentz frequency. Quantity $l=(n_d^2+2)/3$ is the so-called
local-field enhancement factor originating from the polarization of
the background dielectric with refractive index $n_d$ by the
embedded active particles \cite{Crenshaw}.

Further, we numerically solve Eqs. (\ref{dPdtau})--(\ref{Maxdl})
using the finite-difference time-domain (FDTD) approach
\cite{Taflove}. The numerical scheme is the same as in Ref.
\cite{Novit2009}. This scheme allows us to solve the Maxwell-Bloch
equations without separation of the field into forward and backward
waves. Extraction of the transmitted and reflected radiation as well
as setting the boundary conditions with launch of the incident pulse
are performed with the total-field / scattered-field (TF/SF) method.
We also apply the perfectly matched layer (PML) absorbing conditions
to eliminate the nonphysical reflections from the edges of the
calculation region \cite{Berenger,Anantha}.

We consider a simple model of disorder, in which the density of
active atoms (or, equivalently, the strength of light-matter
coupling) experiences periodical random variations (Fig.
\ref{fig0}). In other words, the Lorentz frequency at some point $z$
of the medium is given by
\begin{eqnarray}
\omega_L(z) = \omega^0_L [1+2 r (\zeta(z)-0.5)], \label{randvar}
\end{eqnarray}
where $\omega^0_L$ is the constant determined by the average density
of two-level particles, $\zeta(z)$ is the random number uniformly
distributed in the range $[0; 1]$, and $r$ is the parameter of
disorder strength. For $r=0$, we have perfectly ordered case with
the uniform Lorentz frequency $\omega_L (z)=\omega^0_L$, whereas the
case of maximal disorder ($r=1$) means that $\omega_L (z)$ changes
in the range $[0; 2 \omega^0_L]$. In this paper, we study dependence
of pulse propagation on the four main parameters considered in the
framework of this model: (i) the disorder parameter $r$ itself, (ii)
the average Lorentz frequency $\omega^0_L$ (average atom density),
(iii) the total thickness of the disordered medium $L$, and (iv) the
periodicity of random variations of density of active particles
$\delta L$.

The SIT effect was experimentally observed in a number of systems
\cite{Poluektov}, both gaseous (vapors of alkali metals and
rare-earth-metal ions, molecular gases) and solid (semiconductors,
doped crystals). As specific and more recent examples, we mention
SIT in erbium-doped fibers \cite{Nakazawa} and quantum-dot
waveguides \cite{Schneider}, which well correspond to our
one-dimensional problem. As to realization of disorder, Eq.
(\ref{randvar}), we can speculate that the necessary density
variations can be provided in gaseous atomic setups with the optical
trapping technique \cite{Schaetz}, while in solid-state systems the
concentration variation of active particles (e.g., quantum dots) can
result from the proper variations of synthesis conditions, such as
temperature \cite{Thomassen}. On the other hand, one can expect that
further theoretical studies will significantly lower requirements
for the experimental samples to observe the effects of disorder.

In further calculations, we assume, for simplicity, the exact
resonance ($\delta=0$) and vacuum as the background dielectric
($n_d=1$). The relaxation times $T_1=1$ ns and $T_2=0.1$ ns
correspond to both semiconductor quantum dots (artificial atoms) and
rare-earth ions as active two-level particles. It is important for
us here that both relaxation times are much greater than the pulse
duration $t_p=50$ fs; i.e., we operate in the coherent regime of
light-matter interaction. The Lorentz frequency $\omega^0_L$ is
considered in the range from $10^{11}$ to $10^{12}$ s$^{-1}$, which
is strong enough to expect substantial effects of light-matter
coupling on the distances up to $L~1000 \lambda$, where
$\lambda=0.8$ $\mu$m is the central frequency of incident light
pulses. One can expect that analogous effects can be obtained for
lower light-matter couplings, if we take longer paths of pulse
propagation, for the influence of nonlinearity and disorder to have
enough distance to accumulate. We assume that the pulses have a
Gaussian envelope $\Omega=\Omega_p \exp{(-t^2/2t_p^2)}$ with the
peak Rabi frequency $\Omega_p$ measured in the units of
$\Omega_0=\lambda/\sqrt{2 \pi} c t_p$, which corresponds to the
pulse area of $2 \pi$. Such pulses form the so-called $2 \pi$
solitons, which are the main feature of SIT. In this paper, we
consider propagation of $3 \pi$ pulses (i.e., pulses with the
initial amplitude $\Omega_p=1.5 \Omega_0$), which eventually
transform into standard $2 \pi$ solitons when propagating in the
medium \cite{Novit2011}. We prefer to work with such pulses, since
they have higher intensity and, hence, move faster, making them
favorable from the computational viewpoint.

\section{\label{result}Results of calculations}

\begin{figure}[t!]
\includegraphics[scale=0.5, clip=]{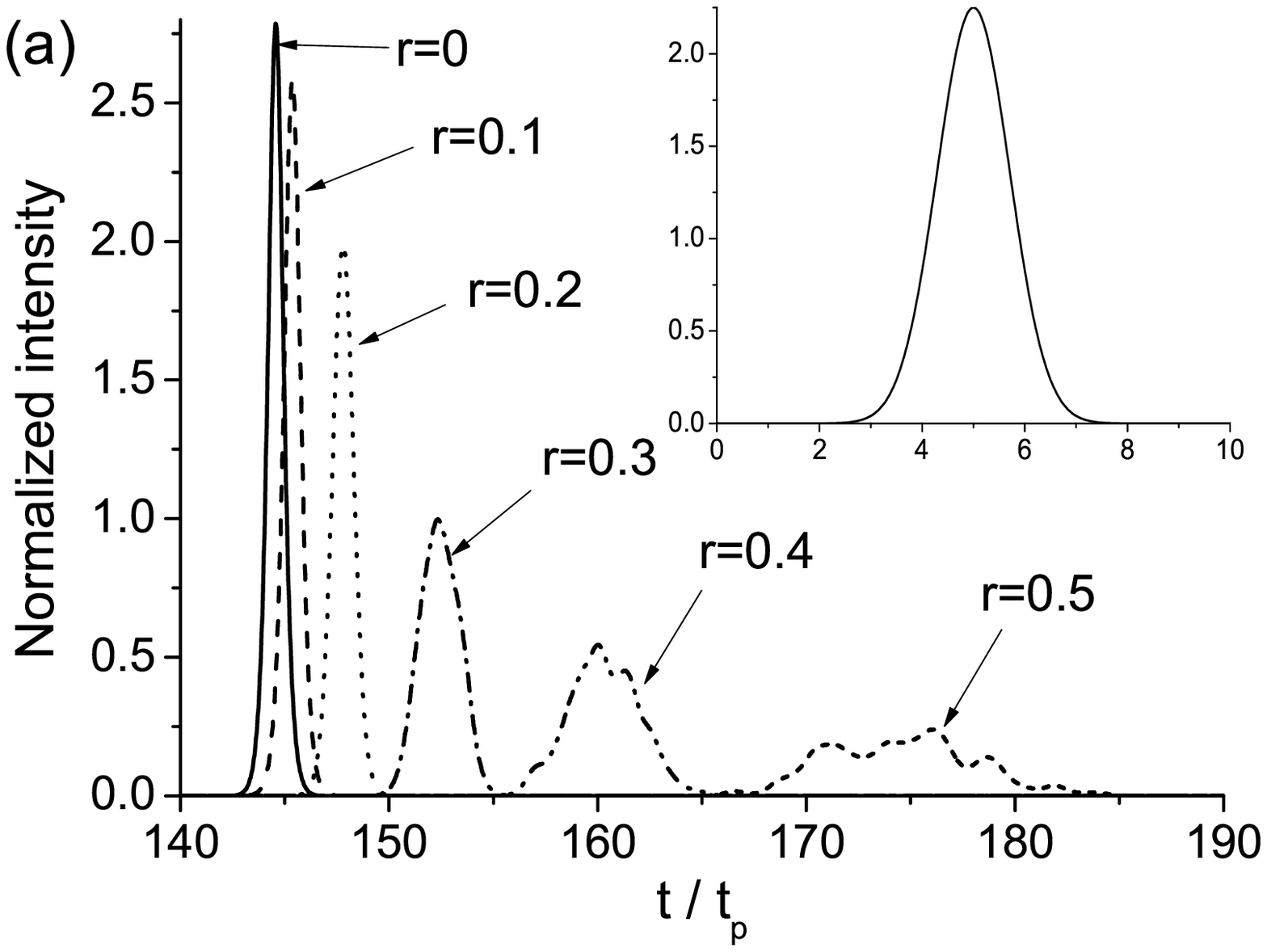}
\includegraphics[scale=0.45, clip=]{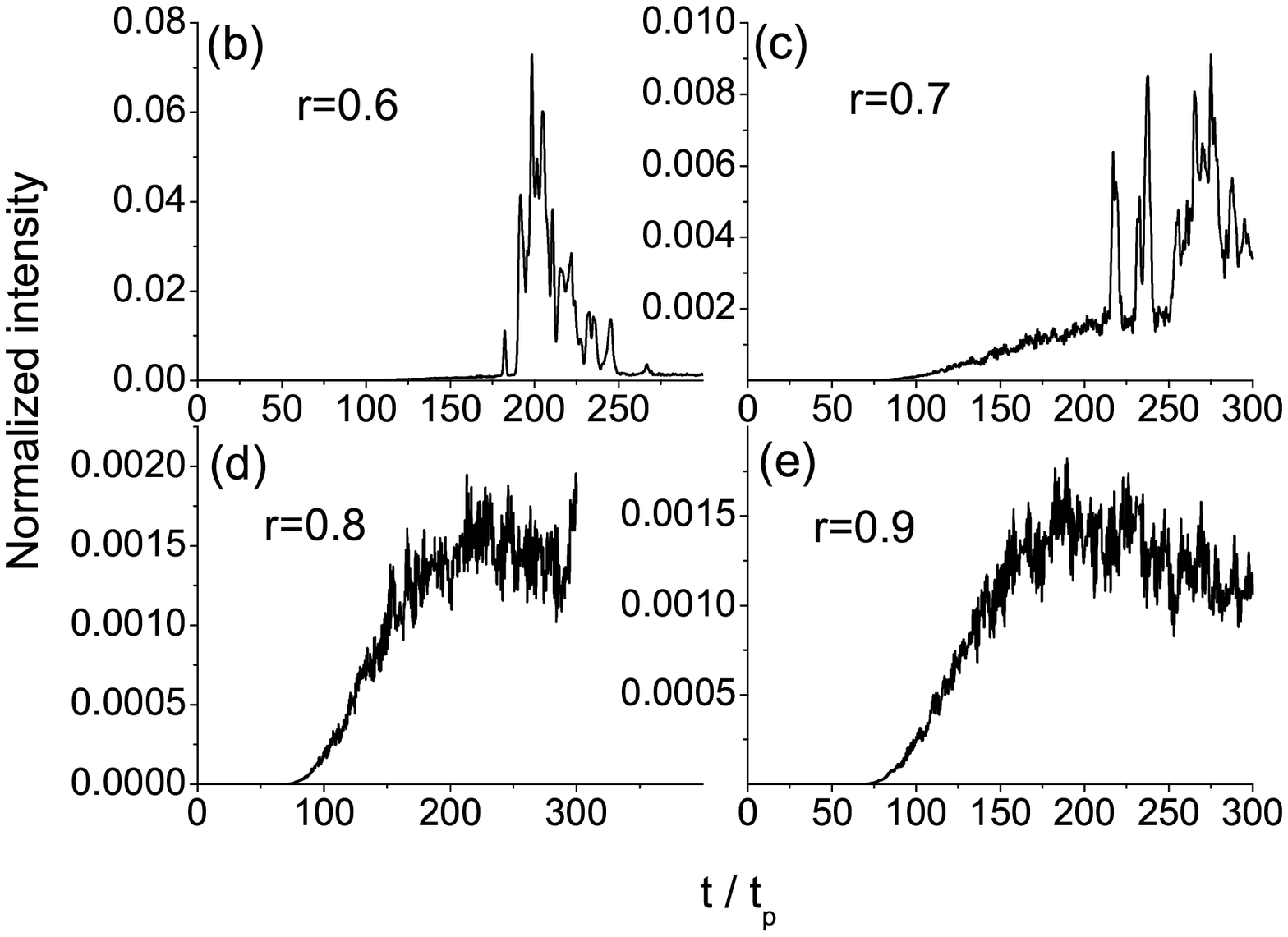}
\caption{\label{fig1} Profiles of transmitted intensity for
different values of the disorder parameter $r$ averaged over $100$
realizations. Other parameters: $L=1000 \lambda$,
$\omega^0_L=10^{12}$ s$^{-1}$, $\delta L=\lambda/4$. The inset shows
the profile of the incident $3 \pi$ pulse.}
\end{figure}

\begin{figure}[t!]
\includegraphics[scale=0.49, clip=]{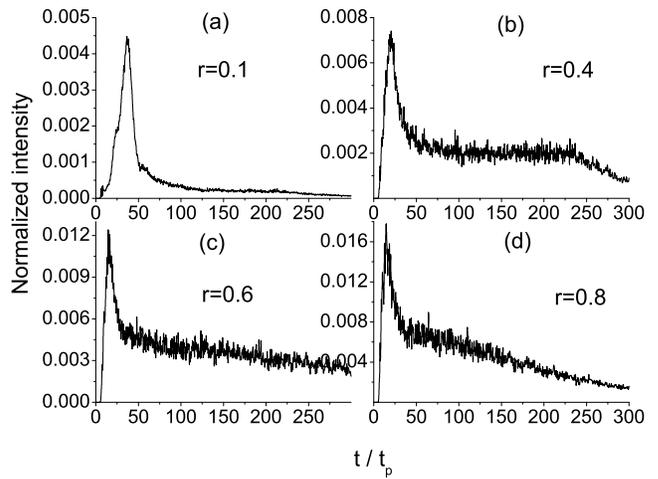} \caption{\label{fig2} Profiles of
reflected intensity for different values of the disorder parameter
$r$ averaged over $100$ realizations. Other parameters are the same
as in Fig. \ref{fig1}.}
\end{figure}

\begin{figure*}[t!]
{\includegraphics[scale=1., clip=]{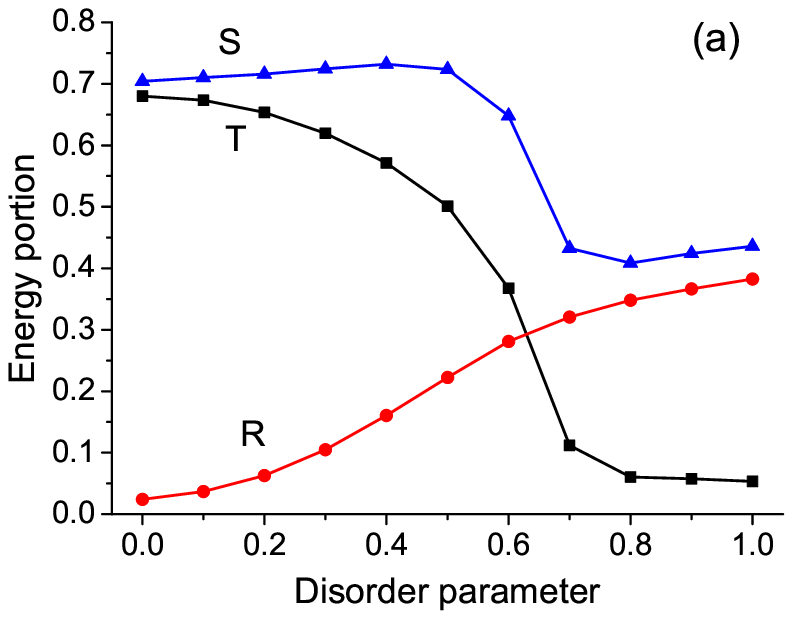}\includegraphics[scale=1.,
clip=]{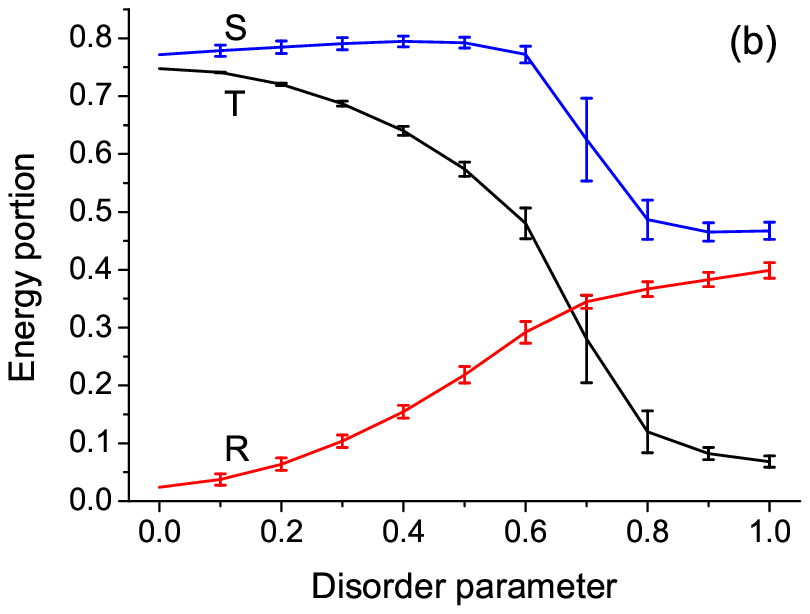}} \caption{\label{fig3} (Color online) Average output
energy of transmitted (T) and reflected (R) light as well as their
sum (S), depending on the disorder parameter $r$. Energy averaged
over $100$ realizations was calculated for the time intervals (a)
$300 t_p$ and (b) $500 t_p$ and was normalized on the input energy.
The parameters are the same as in Fig. \ref{fig1}. The error bars in
panel (b) show the unbiased standard deviations for the
corresponding average values.}
\end{figure*}

\begin{figure*}[t!]
{\includegraphics[scale=0.5, clip=]{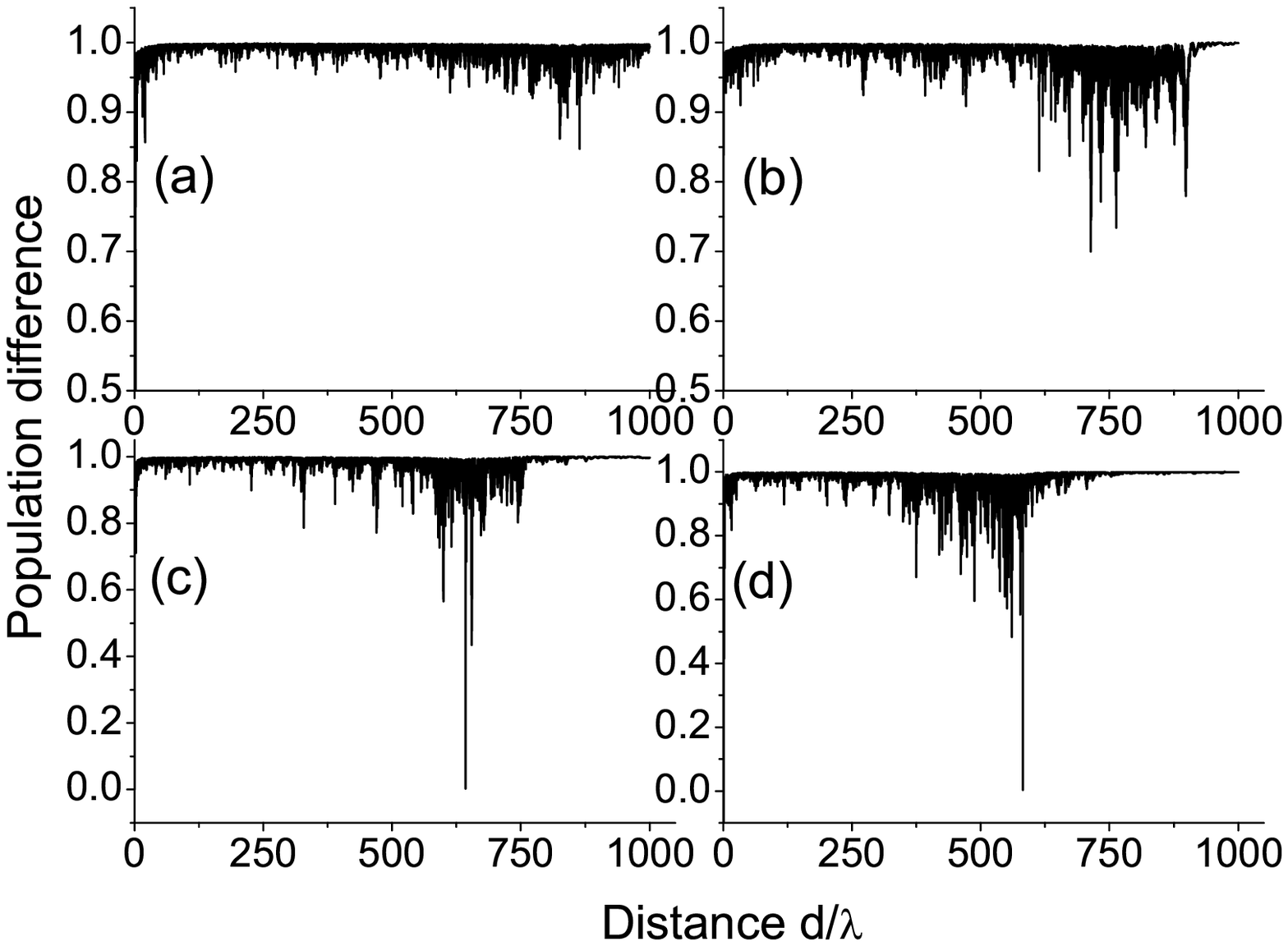}\includegraphics[scale=0.9,
clip=]{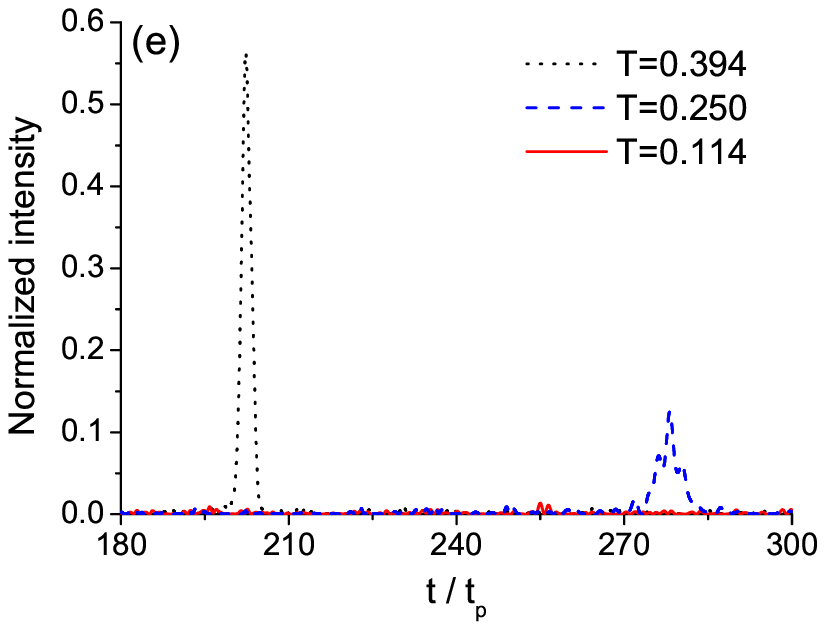}} \caption{\label{fig4} (Color online) [(a)--(d)]
Distribution of population difference along the medium for different
realizations and disorder parameters: (a) $r=0.7$, $T=0.297$,
$R=0.309$; (b) $r=0.7$, $T=0.086$, $R=0.358$; (c) $r=0.8$,
$T=0.059$, $R=0.360$; and (d) $r=1$, $T=0.050$, $R=0.375$. The
distributions are plotted for the time instants [(a), (b)] $350 t_p$
and [(c), (d)] $300 t_p$. (e) The examples of transmitted pulse for
different realizations at $r=0.7$. The parameters are the same as in
Fig. \ref{fig1}.}
\end{figure*}

\begin{figure}[t!]
\includegraphics[scale=1., clip=]{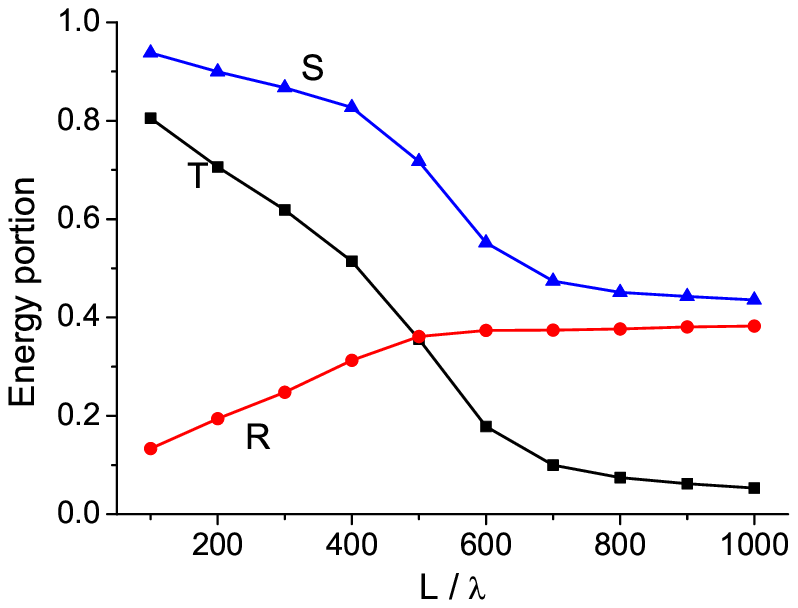} \caption{\label{fig5}
(Color online) Output energy of transmitted (T) and reflected (R)
light as well as their sum (S) depending on the medium thickness
$L$. The parameters used are $r=1$, $\omega^0_L=10^{12}$ s$^{-1}$,
$\delta L=\lambda/4$. Energy averaged over $100$ realizations was
calculated for the time interval $300 t_p$ and was normalized on the
input energy.}
\end{figure}

Let us start with the analysis of $3 \pi$ pulse transmission through
the disordered resonant medium of thickness $L=1000 \lambda$ and
coupling parameter $\omega^0_L=10^{12}$ s$^{-1}$. We assume that the
periodicity of random density variations is $\delta L=\lambda/4$;
i.e., $\omega_L$ randomly changes every quarter wavelength. Figure
\ref{fig1} shows the average profiles of pulses transmitted through
the medium for different values of disorder parameter $r$. Here and
throughout the rest of the paper, averaging is performed over
$N=100$ realizations. At $r=0$, we have usual $2 \pi$ SIT soliton
slightly compressed in comparison with the incident pulse. As $r$
grows, the transmitted pulse gets wider and less powerful and, as a
result, its speed dramatically drops. Nevertheless, we can say that
the average pulse still resembles the usual SIT pulse for $r \leq
0.3$. For stronger disorder, the average profiles become more
distorted, so that we cannot talk about a single transmitted pulse
already at $r=0.5$. At even stronger disorders, there is no pulse at
all in the output, but only low-intensity quasistationary radiation
[Figs. \ref{fig1}(d) and \ref{fig1}(e)] corresponding to the gradual
exit of light after wandering inside the disordered medium. The same
dependence is seen for the reflected intensity profiles presented in
Fig. \ref{fig2}. These profiles have the shape of relatively
low-intensity peaks with subsequent steep decay, which can be
attributed to light slowly leaving the medium. Note that we do not
see particularly strong enhancement of the peak level of reflection
with increasing disorder parameter $r$.

The average generalized characteristics of light interaction with
the disordered resonant medium are shown in Fig. \ref{fig3}(a). In
particular, we demonstrate the output portion of energy, leaving the
medium both in the form of transmitted and reflected radiation for
the time interval of $300 t_p$. We see the gradual increase of
reflection in full accordance with the behavior expected for
disordered media. The curve for transmission has a certain
peculiarity: After smooth attenuation corresponding to the slight
increase in total output at low disorders, there is an abrupt drop
in transmission at $r > 0.5$ with subsequent almost constant level
of transmitted energy. This drop is also clearly seen on the curve
of total output. In part, this can be explained as a result of
dramatic slowing down of the SIT soliton with decrease in its
intensity, so that it is still moving inside the medium at the end
of time interval considered ($300 t_p$). Another reason for this
drop is the transition from the regime of pulse transmission [very
weak and very slow remainder of the pulse can be seen in Fig.
\ref{fig1}(b) for $r=0.6$] to the regime of smooth rise of the
transmitted intensity without any pulse at the exit [see Fig.
\ref{fig1}(c) for $r=0.7$]. In other words, we see \textit{the
transition from SIT to light localization}. This conclusion is
corroborated in Fig. \ref{fig3}(b), where larger time interval ($500
t_p$) is used for energy calculations. We see that the output energy
is obviously larger in comparison with that in Fig. \ref{fig3}(a),
since all the slow solitons have now enough time to pass the medium.
Nevertheless, the overall trend remains the same. In particular,
there is still the abrupt drop of the transmitted and total output
energies, which is the evidence of the localization threshold. This
threshold is slightly higher for larger time interval than for
shorter. It is unlikely that this threshold will change
significantly with further increase of time interval. Therefore, in
the rest of the paper, we use the time interval of $300 t_p$ to
calculate average energetic characteristics, since it is long enough
to give the correct representation of the behavior of the system.

Notice, that the total absence of pulse in the average transmission
profile does not mean that there is no realizations with pulse in
the output; these realizations just become less frequent. This is
corroborated by the direct comparison of the number of realizations
with different transmission. For small disorder ($r \leq 0.3$), all
realizations have similar transmission close to that of the SIT
soliton in ordered resonant medium (in the range from $0.6$ to $0.7$
for the time interval of $300 t_p$). Increasing disorder results in
growing number of realizations with low transmission. At $r=0.6$,
most realizations have transmission of $0.3-0.4$, while already for
$r=0.7$ more than a half of realizations have transmission of the
order of $0.1$. This statistics correspond to the drop in Fig.
\ref{fig3}(a). Nevertheless, at $r=0.7$, a large portion of
realizations still has relatively large transmission up to $0.4$
with a low-intensity pulse in the output. We illustrate this with
several examples of transmitted pulses for different realizations
[Fig. \ref{fig4}(e)]: One can see the realizations with a sharp
pulse ($T=0.394$), without pulse at all ($T=0.114$), and the
intermediate case ($T=0.250$). We emphasize that this diversity of
realizations is characteristic for the disorder strengths
corresponding to transition from SIT to localization. For lower and
higher disorder parameters, we have mostly one of two possibilities:
either strong pulse, or strong localization. This change in
statistics of realizations can be illustrated in a different way.
The unbiased standard deviation shown with error bars in Fig.
\ref{fig3}(b) is calculated using the well-known formula
\cite{Brandt} $s=\sqrt{\sum(E_i - \overline{E})^2 / (N-1)}$, where
$E_i$ stands for transmitted, reflected, or total output energy in a
$i$th realization, $\overline{E}$ is the corresponding average
value, and $N=100$ is the number of realizations. It is seen that
the standard deviation is the largest for the transition between the
regimes of SIT (small $r$) and localization (large $r$). This fact
reflects the diversity of realizations discussed above. The same
regularity is characteristic for other plots of energy in this
paper. We should also note that error bars in Fig. \ref{fig3}(b)
confirm the statistical significance of the results obtained.

To get an idea of the fate of light in the medium, it is worth
considering population difference distributions calculated for a
couple of realizations with high and low transmission as given in
Figs. \ref{fig4}(a) and \ref{fig4}(b). As expected, the low output
is associated with larger residual excitation of the medium in
comparison to the high-transmission case. For $r \geq 0.8$,
practically all realizations are characterized by transmission less
than $0.1$ that corresponds to larger portion of light energy left
in the medium as evidenced by the residual population differences
plotted in Figs. \ref{fig4}(c) and \ref{fig4}(d). These examples of
realizations also show that growing disorder results not only in
increase of energy localized inside the medium, but also in the
shift of residual excitation to the entrance of the medium; i.e., a
larger portion of light is trapped closer to the input of strongly
disordered medium.

\begin{figure}[t!]
\includegraphics[scale=1., clip=]{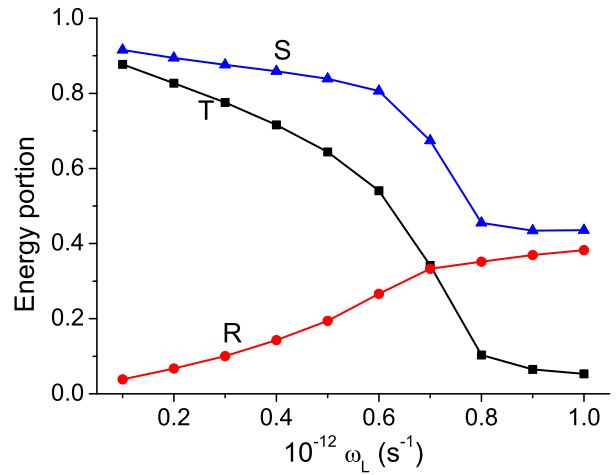} \caption{\label{fig6}
(Color online) Output energy of transmitted (T) and reflected (R)
light as well as their sum (S) depending on the average light-matter
coupling $\omega^0_L$. The parameters used are $r=1$, $L=1000
\lambda$ s$^{-1}$, $\delta L=\lambda/4$. Energy averaged over $100$
realizations was calculated for the time interval $300 t_p$ and was
normalized on the input energy.}
\end{figure}

\begin{figure}[t!]
\includegraphics[scale=1., clip=]{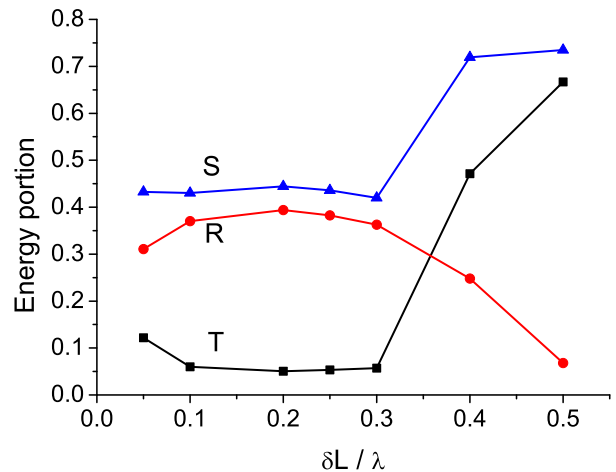} \caption{\label{fig7}
(Color online) Output energy of transmitted (T) and reflected (R)
light as well as their sum (S) depending on the period of random
variations $\delta L$. The parameters used are $r=1$, $L=1000
\lambda$ s$^{-1}$, $\omega^0_L=10^{12}$ s$^{-1}$. Energy averaged
over $100$ realizations was calculated for the time interval $300
t_p$ and was normalized on the input energy.}
\end{figure}

The next issue is the dependence of localization effect on another
parameter -- the medium thickness $L$. The energy curves calculated
for the time interval $300 t_p$ are shown in Fig. \ref{fig5}. The
transmission here monotonously decreases, while reflection
demonstrates clear saturation with distance. This result was
obtained for the quite large light-matter coupling
$\omega^0_L=10^{12}$ s$^{-1}$, but similar dependence is expected
for lower couplings and larger distances. The importance of the
parameter $\omega^0_L$ (or, equivalently, average density of
two-level atoms) is illustrated with Fig. \ref{fig6}. It
demonstrates again the drop in total output at $\omega^0_L > 6
\times 10^{11}$ s$^{-1}$; i.e., there is the threshold value of
light-matter interaction, which governs the transition to the
localization regime characterized by the absence of the transmitted
pulse and almost constant output approximately equal to $0.5$; i.e.,
the saturation of the output is present as well.

Finally, we should study the influence of the period of random
atom-density variations $\delta L$ on pulse localization. The
results of energy calculations shown in Fig. \ref{fig7} indicate
that the optimal localization occurs for $\delta L$ in the range
from $0.1 \lambda$ to $0.3 \lambda$. At larger periods, transmission
sharply increases, while reflection decreases that indicates
transition to the SIT regime. The same is likely to be true for
$\delta L < 0.1 \lambda$, though the rise of transmission is not so
pronounced in this case.

\section{\label{concl}Conclusion}

In conclusion, the concept of disordered resonant medium and the
simple model of disorder introduced in this paper substantially
broadens the scope of nonlinear disordered photonics. We have shown
that increase of disorder level results in transition between the
regimes of self-induced transparency and light localization in a
medium with randomly changing density of active particles. This
effect is strongly influenced not only by the disorder parameter,
but also by the average atom density (or, equivalently, the average
light-matter coupling), so that one can talk about the localization
threshold for these parameters. There is also an optimal range of
random variations period for the localization to occur, which
implies that atom density should change on the distances of the
order of the quarter wavelength.

Although SIT and similar effects can be naturally analyzed in one
dimension, it would be interesting to consider the problem of two-
and three-dimensional disordered resonant media. Another possible
direction is to go beyond the two-level approximation to study, for
example, such effects as electromagnetically induced transparency in
three- and four-level disordered systems. Thus, incorporation of
disorder into the nonlinear resonant media opens interesting avenues
in semiclassical physics of light-matter interaction.

\acknowledgements{The author acknowledges financial support from the
Russian Science Foundation (Project No. 17-72-10098).}


\begin{thebibliography}{0}
\bibitem{Wiersma} D. S. Wiersma, {Nat. Photon.} {\bf7}, 188 (2013).
\bibitem{Segev} M. Segev, Y. Silberberg, and D. N. Christodoulides, {Nat. Photon.} {\bf7}, 197 (2013).
\bibitem{Conti2007} C. Conti, L. Angelani, and G. Ruocco, {\pra} {\bf75}, 053827 (2007).
\bibitem{Folli13} V. Folli, K. Gallo, and C. Conti, {\ol} {\bf38}, 5276 (2013).
\bibitem{Folli12} V. Folli and C. Conti, {\ol} {\bf37}, 332 (2012).
\bibitem{JovicOL} D. M. Jovi\'{c}, C. Denz, and M. R. Beli\'{c}, {\ol} {\bf37}, 4455 (2012).
\bibitem{JovicPRA} D. M. Jovi\'{c}, M. R. Beli\'{c}, and C. Denz, {\pra} {\bf85}, 031801(R) (2012).
\bibitem{Cherroret} N. Cherroret, B. Vermersch, J. C. Garreau, and D. Delande, {\prl} {\bf112}, 170603 (2014).
\bibitem{Schwiete} G. Schwiete and A. M. Finkel'stein, {\prl} {\bf104}, 103904 (2010).
\bibitem{Muskens} O. L. Muskens, P. Venn, T. van der Beek, and T. Wellens, {\prl} {\bf108}, 223906 (2012).
\bibitem{Skokos} Ch. Skokos, I. Gkolias, and S. Flach, {\prl} {\bf111}, 064101 (2013).
\bibitem{Leykam} D. Leykam, S. Flach, O. Bahat-Treidel, and A. S. Desyatnikov, {\prb} {\bf88}, 224203 (2013).
\bibitem{Conti2012} C. Conti, {\pra} {\bf86}, 061801(R) (2012).
\bibitem{Novitsky2014} D. V. Novitsky, {\josab} {\bf31}, 1282 (2014).
\bibitem{Novitsky2015} D. V. Novitsky, {\oc} {\bf353}, 56 (2015).
\bibitem{McCall} S. L. McCall and E. L. Hahn, {Phys. Rev.} {\bf183}, 457 (1969).
\bibitem{Poluektov} I. A. Poluektov, Yu. M. Popov, and V.S. Roitberg, {Sov. Phys. Usp.} {\bf17}, 673 (1975).
\bibitem{Hopf} F. A. Hopf, C. M. Bowden, and W. H. Louisell, {\pra} {\bf29}, 2591 (1984).
\bibitem{Afan98} A. A. Afanas'ev, R. A. Vlasov, N. B. Gubar, and V. M. Volkov, {\josab} {\bf15}, 1160 (1998).
\bibitem{Ponomarenko} S. A. Ponomarenko and S. Haghgoo, {\pra} {\bf82}, 051801 (2010).
\bibitem{Novitsky2017} D. V. Novitsky, {\pra} {\bf95}, 053846 (2017).
\bibitem{Pakhomov} A. V. Pakhomov, R. M. Arkhipov, I. V. Babushkin, M. V. Arkhipov, Yu. A. Tolmachev, and N. N. Rosanov, {\pra} {\bf95}, 013804 (2017).
\bibitem{Arkhipov2016} R. M. Arkhipov, M. V. Arkhipov, I. Babushkin, A. Demircan, U. Morgner, and N. N. Rosanov, {\ol} {\bf41}, 4983 (2016).
\bibitem{Arkhipov2017} R. M. Arkhipov, A. V. Pakhomov, M. V. Arkhipov, I. Babushkin, A. Demircan, U. Morgner, and N. N. Rosanov, {Sci. Rep.} {\bf7}, 12467 (2017).
\bibitem{Folli11} V. Folli and C. Conti, {\ol} {\bf36}, 2830 (2011).
\bibitem{Novit2011} D. V. Novitsky, {\pra} {\bf84}, 013817 (2011).
\bibitem{Crenshaw} M. E. Crenshaw, {\pra} {\bf78}, 053827 (2008).
\bibitem{Taflove} A. Taflove and S. C. Hagness, \textit{Computational Electrodynamics: The
Finite-Difference Time-Domain Method}, 2nd ed. (Artech House,
Boston, 2000).
\bibitem{Novit2009} D. V. Novitsky, {\pra} {\bf79}, 023828 (2009).
\bibitem{Berenger} J. P. Berenger, {J. Comput. Phys.} {\bf114}, 185 (1994).
\bibitem{Anantha} V. Anantha and A. Taflove, {IEEE Trans. Ant. Prop.} {\bf50}, 1337 (2002).
\bibitem{Nakazawa} M. Nakazawa, Y. Kimura, K. Kurokawa, and K. Suzuki, {\prb} {\bf45}, R23 (1992).
\bibitem{Schneider} S. Schneider, P. Borri, W. Langbein, U. Woggon, J. F\"{o}rstner, A. Knorr, R. L. Sellin, D. Ouyang, and D. Bimberg, {\apl} {\bf83}, 3668 (2003).
\bibitem{Schaetz} T. Schaetz, {J. Phys. B} {\bf50}, 102001 (2017).
\bibitem{Thomassen} S. F. Thomassen, T. W. Reenaas, and B. O. Fimland, {J. Cryst. Growth} {\bf323}, 223 (2011).
\bibitem{Brandt} S. Brandt, \textit{Data Analysis: Statistical and Computational Methods for Scientists and Engineers},
4th ed. (Springer, New York, 2014).
\end{thebibliography}
\end{document}